\documentclass[twocolumn,letter]{jpsj2}
\usepackage{bm}

\title{
Interference Effect in Multi-level Transport through a Quantum Dot
}

\author{Hisashi~\textsc{Aikawa}\thanks{E-mail:
astar@issp.u-tokyo.ac.jp}, Kensuke~\textsc{Kobayashi},
Akira~\textsc{Sano}, Shingo~\textsc{Katsumoto}, and
Yasuhiro~\textsc{Iye}}

\inst{Institute for Solid State Physics, University of Tokyo, 5-1-5
Kashiwanoha, Chiba 277-8581}

\recdate{}

\abst{We present experimental results and a model to solve the problem
of ``in-phase Coulomb peaks" observed in transport through a quantum
dot.  In a marginal region between Coulomb-blockade and open-dot, we
have observed Fano-type interference through two energy levels inside
the dot, which manifest themselves in two overlapped
Coulomb-diamond-like structures in the excitation spectrum.  One of
the two levels is strongly coupled to the leads and the phase of
traversing electrons is locked to it.  We have detected the phase
change at the vertices and the centers of the larger diamonds through
the sign of the Fano's asymmetric parameters supporting the above
deduction.}

\kword{quantum dot, coherent transport, phase shift, Fano effect}

\begin{document}
\sloppy
\maketitle 

A particle scattering experiment gives us information of the scatterer
through the cross section and the phase shift.  Electric conduction
through a quantum dot (QD) can be viewed as scattering process and the
phase shift and the transmission amplitude furnish with useful pieces
of information on the electronic states in the QD.  One of the ways to
extract the phase shift is a double-slit experiment. Such an
experiment was first carried out by Yacoby {\it et al.} using an
Aharonov-Bohm (AB) ring with a QD~\cite{Yacoby1995PRL}. The AB phase
shift was found to jump by $\pi$ just at every Coulomb peak due to the
two-terminal configuration~\cite{Yeyati1995PRB} of the sample. The
experiment adopting four-terminal
configuration~\cite{Schuster1997Nature} demonstrated Breit-Wigner-type
gradual phase change by $\pi$ at each peak rather than a jump.  A
surprising finding, however, was that the phase additionally jumps by
$\pi$ between adjacent peaks and the phase returns to the starting
value after a single cycle of Coulomb oscillation. This ``phase lapse"
problem can be divided into two: (A) Why do the additional phase
changes of $\pi$ appear between Coulomb peaks? (B) Why do they appear
as ``jumps" even in the four-terminal geometry?  Here we call (A) the
problem of ``in-phase Coulomb peaks", which we treat in this Letter.
The same phenomenon has been observed as peaks with the same sign of
Fano's asymmetric parameter~\cite{Kobayashi2002PRL}.

A number of theoretical models have been proposed to explain this
problem~\cite{Yeyati1995PRB,Bruder1996PRL,Oreg1997PRB,Xu1998PRB,Deo1998,RyuCho,Kang,Wu1998PRL,Lee1999PRL,Kim,Kim2003,Aharony2002PRB,Aharony2003}.
All of them seem to reproduce the main experimental features, and it
is a task of further experiments to find out which of them or another
mechanism is appropriate.  We have experimentally proven~\cite{T-Fano}
that the in-phase Coulomb peaks appear in a quantum wire with a
side-coupled QD, which rules out a certain class of models that base
themselves on the specific geometry of the AB resonator.

Here we introduce a possible model based on a simple single-electron
model motivated by a numerical simulation by Nakanishi {\it et
al.}~\cite{Nakanishi}. A quantum state inside a rectangular QD is
labeled by a wave number $\bm{k}_{ln}=(\frac{2\pi}{L_x}l,
\frac{2\pi}{L_y}n)$ ($L_x > L_y$), where the indices $l$ and $n$ are
positive integers.  Consider the situation where two leads are
attached to the QD via tunnel barriers along the $x$-direction.  In
the transport, the single-electron levels indexed by $(l,n)$ appear as
Coulomb peaks.  For a given $l$, the $(l,1)$ state among the series
$(l,n)$ most strongly couples with the leads because the kinetic
energy along the $x$-direction is highest, which makes the effective
barrier height lowest. Hence the QD states are classified into a small
number of strong coupling states (SCSs) and the others (weak coupling
states, WCSs).  Note that the SCSs correspond to classical
trajectories that directly cross the dot while the WCSs correspond to
those with many bounces along $y$.

The existence of such SCSs in more realistic dot potentials with leads
is reported in numerical
simulations~\cite{Silvestrov2000PRL,Yeyati2000PRB}.  The strong
couplings occur mainly through a connection of the inlet and the
outlet by a large amplitude part of the wave function, which has a
corresponding classical trajectory (a kind of
``scar")~\cite{stopa,YHKim2002}.  Hence the situation in the general
models is essentially the same as that in the simple symmetric model
besides the phase shifts as we see below.

Let us go back to the rectangular model and introduce some distortion
(or disorder) expressed by a perturbation potential $V$, which causes
an intermixing of the $x$- and $y$-motions.  The wave function of a
WCS after the mixing is expressed in the first order as
\begin{equation}
\psi_j \approx \psi_j^0+\psi_N\frac{\langle
\psi_j^0|V|\psi_N\rangle}{E_j^0-E_N},
\label{perturbation}
\end{equation}
where $N$ is the index of SCS closest to the energy level $E_j^0$ of
the unperturbed state $\psi_j^0$.  The contributions from other WCSs
can be ignored from the viewpoint of transport.  Because $\psi_N$ has
much stronger coupling with the lead states, the transport through
$\psi_j$ is dominated by the second term in the right hand side of
Eq.~(\ref{perturbation}).  This leads to a series of in-phase peaks
and the phase changes only when the closest SCS takes over due to the
energy denominator in Eq.~(\ref{perturbation}). Note that this
discussion is applicable to any dot shape if
$\langle\psi_j^0|V|\psi_N\rangle$ is finite.  It should also be noted
that this model only explains why there are series of in phase Coulomb
peaks.  The problem how the phase goes back to the initial value ({\it
i.e.,} problem (B)) between the peaks is out of our scope here.

Because an SCS has a large level broadening due to the strong
coupling, the one closest to $E_{\rm F}$ gives ``background"
conduction with constant phase shift like a continuum to that through
the WCS in question.  In such systems (a discrete energy level plus
the continuum), Fano showed that there appears a characteristic
distortion in the transition probability (the Fano
effect)~\cite{Fano1961PR}.  In the case of transport through a quantum
dot, the lineshape of the conductance $G$ to the gate voltage $V_{\rm
g}$ is given as,
\begin{equation}
	G (V_{\rm g}) = A {(\epsilon + q)^2 \over \epsilon^2 + 1},
\label{eqFano}
\end{equation}
where $\epsilon \equiv {\alpha (V_{\rm g} - V_{\rm res}) / (\Gamma
/2)}$, $A$ is the amplitude, $\alpha$, gate voltage-energy conversion
factor, $V_{\rm res}$, resonance position, $\Gamma$, width of the
resonance, and $q$, Fano's asymmetric parameter.  The parallel
conduction in the QD should thus lead to the Fano-type
interference~\cite{Nakanishi}.  This Fano-type distortions in
transmission spectra through single quantum dots without external
interference circuit have been
reported~\cite{Gores,Zacharia,Clerk2000PRL,Fuhner2003CM}, though no
explanation has been given to the origin of parallel conduction.  In
Ref.~17, we have demonstrated that Fano effect can be utilized for the
detection of phase shift variation without AB geometry.  This can be
easily unerstood by considering that the origin of the Fano distortion
is the rapid variation of the phase shift by $\pi$ around the
resonance, {\it i.e.}, rapid change of the interference sign from
constructive to destructive (or vice versa).  Hence the direction and
the degree of the distortion, which is represented by Fano's parameter
$q$, is sensitive to the variation of the phase shift.  Therefore
investigation of the single-dot Fano effect should be the touchstone
of the above model to account for ``in phase Coulomb peaks".  Problem
(B) cannot be treated here because the Fano effect appears only around
the resonances.

In this Letter, we report systematic experiments on the single-dot
Fano interference in a multi-level transport regime. We have observed
clear trace of SCSs and changes of sign of the Fano parameter in
accordance with the inference. The present result provides a simple
explanation for the long-standing ``in phase Coulomb peaks" puzzle.

\begin{figure}[tbp]
\begin{center}
\includegraphics[width=\linewidth]{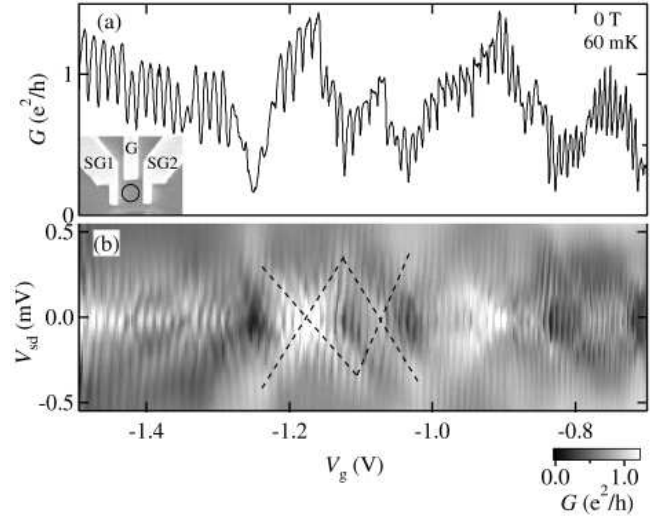}
\caption{(a) QD conductance $G$ vs gate voltage $V_{\rm g}$ in the
multi-level transport regime. Rapid oscillations that originate from
the WCSs are superimposed on the slow background oscillation.  The
inset shows a scanning electron micrograph of the sample.  The white
regions are metallic gates that define a QD as illustrated by the
black circle (200~nm diameter). The side gates SG1 and SG2 determine
the coupling to the leads and the center gate G mainly modifies the
electrostatic potential of the QD.  The lower region is etched off
(not well resolved in this picture).  (b) Gray-scale plot of $G$ as a
function of bias voltage $V_{\rm sd}$ and $V_{\rm g}$. Fine
stripe-like features are Coulomb diamonds of WCS. On top of these fine
structures, high conductance (white) regions show systematic change,
leading to the large Coulomb-diamond-like structure (dashed lines).}
\label{Fig1}
\end{center}
\end{figure}

We prepared a QD from 2DEG formed at GaAs/AlGaAs hetero-structure
(sheet carrier density $3.8 \times 10^{15}~{\rm m^{-2}}$,
mobility $80~{\rm m^2/Vs}$) by using electron beam lithography
followed by deposition of metallic gates and wet etching. The
inset of Fig.~\ref{Fig1}(a) shows the gate configuration.  The sample
was cooled in the dilution refrigerator with a base temperature of
30~mK and was measured by standard lock-in techniques in a
two-terminal setup.  In order to enhance the interference effect, we
chose the side gate (SG1 and SG2) voltages to keep the total
conductance around $e^2/h$ where the QD is at the border between the
Coulomb blockade regime and the open-dot regime.

\begin{figure}[tbp]
\begin{center}
\includegraphics[width=\linewidth]{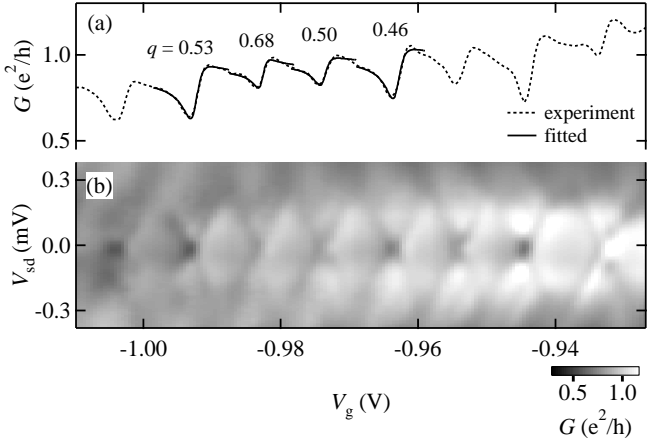}
\caption{Enlarged view of Fig.~\ref{Fig1}. (a) Line shapes of the
Coulomb oscillation show the Fano effect.  Fitting results to
Eq.~(\ref{eqFano}) (solid lines) and the obtained values of $q$ are
also shown. Constant background is assumed for the fitting
process. (b) Conductance as a function of both $V_{\rm g}$ and $V_{\rm
sd}$. The shapes of the diamonds are formed by lower conductance
borders.  The estimated $C_{\rm g}$ and the charging energy are 16~aF
and 3.7~K, respectively, which are reasonable for the device
dimensions.}
\label{Fig2}
\end{center}
\end{figure}

Figure~\ref{Fig1}(a) shows the conductance $G$ of the QD as a function
of $V_{\rm g}$.  A fine oscillation is superposed on a slow background
oscillation (BO).  It appears as a sequence of sharp dips at the
lowest temperature, which is the sign of reversed Coulomb oscillation.
These dips rapidly changes to ordinary Coulomb peaks with decreasing
the total conductance.  In Fig.~\ref{Fig1}(b), we show a gray-scale
plot of $G$ on the plane of $V_{\rm g}$ and the source-drain bias
voltage $V_{\rm sd}$.  The Coulomb diamonds are reversed, {\it i.e.},
appear as high conductance regions (also shown in Fig.~\ref{Fig2}(b)),
and their widths along the $V_{\rm sd}$ axis are small (large) around
the peaks (valleys) of the BO.  As a result, there appear larger
diamond structures in Fig.~\ref{Fig1}(b), one of which is indicated by
dashed lines.

The above observations indicate that the slow BO is due to the
conduction through SCSs. Since they are spatially extended, their
Coulomb peaks are smeared in the present condition.  The modulation of
small Coulomb diamonds by $V_{\rm g}$, {\it i.e.}, the appearance of
larger diamonds can be understood from Eq.~(\ref{perturbation}).
Around a peak of the BO, the second term in Eq.~(\ref{perturbation})
is large due to small energy denominator, resulting in large spatial
size and large effective capacitance $C_{\rm eff}$, hence the
corresponding diamond is small.  In a valley of the BO, the conditions
are opposite and larger diamond appears.  The width of the diamonds
mainly oscillates along $V_{\rm sd}$ and only small gradual changes
occur along $V_{\rm g}$.  This is due to the specific shape of the QD
shown in the inset of Fig.~\ref{Fig1}(a).  Because the point contacts
are attached to the other side of the gate electrode, an elongation of
a state with the mixing from an SCS occurs far from the gate.  Hence
the oscillation in $C_{\rm g}$ (the gate capacitance) is small while
$C_{\rm s}$ and $C_{\rm d}$ (the source and drain capacitances)
largely oscillate.  It should be noted that the width of a diamond is
smaller both along $V_{\rm g}$ and $V_{\rm sd}$ for smaller negative
$V_{\rm g}$.  This is because the effective QD size is reduced by
increasing the negative gate voltage.

Figure~\ref{Fig2} shows a magnification of a part of Fig.~\ref{Fig1}.
Clear Fano distortion is observed and each dip can be fitted by Fano's
asymmetric line shape Eq.~(\ref{eqFano}). This confirms that the
two-level transport model is a good approximation for the present
case.  Figure~\ref{Fig2}(b) shows an expanded view of
Fig.~\ref{Fig1}(b).  Outside the borders of clear inverted Coulomb
diamonds, parallel lines due to excitation to a higher level are
observed around $V_{\rm g} = -0.95$~V.  The Fano distortion is
observed only around $V_{\rm sd}=0$ again in accordance with the
previous result~\cite{Kobayashi2002PRL}.

\begin{figure}[tbp]
\begin{center}
\includegraphics[width=\linewidth]{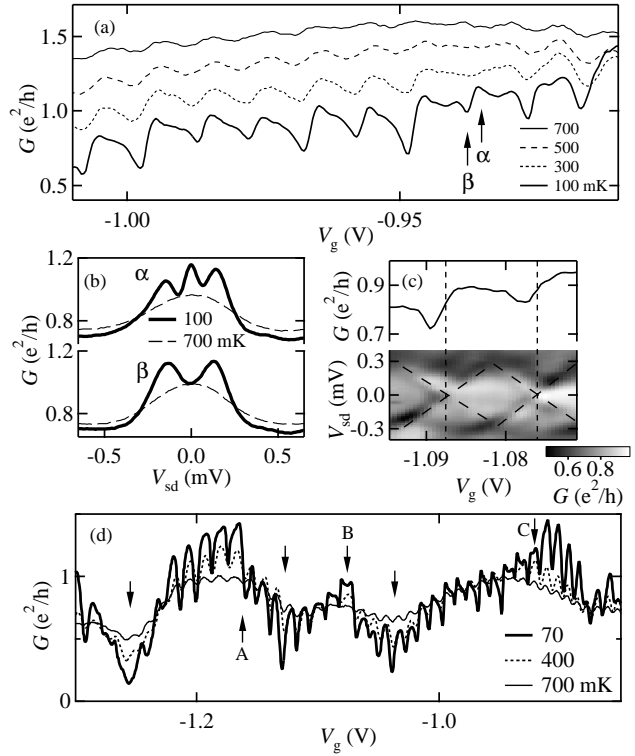}
\caption{(a) Temperature dependence of the Fano interference displayed
with offset for clarity. (b) shows $V_{\rm sd}$ measurement at fixed
$V_{\rm g}$ indicated by arrows in (a). The upper and the lower data
correspond to constructive and destructive interference, respectively.
(c) An example of obtaining a sign of $q$ from the sign of gradient at
a center of resonance (indicated by vertical dotted lines) determined
at a cross point of the Coulomb diamonds (dashed lines in the lower
gray scale plot). (d) Temperature dependence of the SCSs. Arrows
indicate where the sign of $q$ changes (see text).}
\label{Fig3}
\end{center}
\end{figure}

Figure~\ref{Fig3}(a) displays the temperature dependence of the
zero-bias conductance oscillation. The Fano interference are smeared
out at around 500~mK.  We have attributed a smearing of the line shape
of a side-coupled QD to the thermal broadening and quantum
decoherence~\cite{T-Fano}.  Essentially the same discussion is
applicable here.  In Fig.~\ref{Fig3}(b), we present the $V_{\rm sd}$
dependence of $G$ taken at fixed $V_{\rm g}$ as indicated by the
arrows $\alpha$ and $\beta$ in Fig.~\ref{Fig3}(a), where the
interference is constructive and destructive, respectively.  In the
destructive case (the lower in Fig.~\ref{Fig3}(b)), a simple resonance
dip appears at zero-bias while in the constructive case (the upper in
Fig.~\ref{Fig3}(b)), additional side peaks appear at low temperatures.
The interference with an SCS with higher energy may be responsible for
the side peaks though at present we have no concrete idea.
Figure~\ref{Fig3}(d) shows the temperature dependence of the BO, the
amplitude of which simply diminishes with increasing temperature as
expected.

So far we have investigated how multi-level transport appears in the
conductance.  Now we examine the phase shifts at the Coulomb peaks.
The previous works~\cite{Kobayashi2002PRL, Kobayashi2003PRB, T-Fano}
have established that information on the phase shift at a QD can be
obtained from the sign of $q$ and so does the
simulation~\cite{Nakanishi}.  Note that the Breit-Wigner type phase
shift is {\it assumed}, as was experimentally
proven~\cite{Schuster1997Nature}.  Hence the problem we can address in
the present two-terminal configuration is limited to (A) (``in-phase
Coulomb peaks"); We are not concerned here with the phase-jump (B) due
to the phase-locking~\cite{Yeyati1995PRB,Aharony2002PRB,Aharony2003}.
What we expect here are the following.  i) An SCS dominates the
perturbations over a range of WCSs, of which the Coulomb peaks are
in-phase; ii) The sign of $q$ of a Coulomb peak reflects the phase,
provided that the reference phase is unchanged.  The turnover of the
dominant SCS occurs at the valleys of the BO.  From i), therefore, the
Coulomb peaks are in-phase between two adjacent valleys of the BO.
However in the present case, the reference of the Fano interference is
the SCS, whose phase changes by $\pi$ at the corresponding peak of the
BO.  As a result, $q$ should change its sign both at the peaks and the
valleys of the BO.  In the numerical simulation in
Ref.~\cite{Nakanishi}, this behavior is clearly observed.

In order to analyze the data in Fig.~\ref{Fig3}(d) along the above
line, the sign of $q$ should be obtained for all the Coulomb
dips. Unfortunately they are significantly distorted not only by the
Fano effect but also by the BO itself hence if we adopt fitting of
Eq.~(\ref{eqFano}) to obtain $q$, the error bar crosses zero at a
considerable number of dips.  Here, instead, we first determined the
centers of resonances from the Coulomb diamonds as presented by the
dotted vertical lines in Fig.~\ref{Fig3}(c) and obtained the signs of
$q$'s from those of the gradients in conductance at the resonance
position.  This method gives correct sign of $q$ at a peak (dip)
though the absolute value still has a large ambiguity.

We can see the expected behavior in Fig.~\ref{Fig3}(d), where the
zero-crossing points of $q$ are indicated by the arrows. They are
placed at the peaks and the valleys of the BO.  Especially the changes
of the sign at the peaks (labeled as A, B, and C) are clear.

\begin{figure}[tbp]
\begin{center}
\includegraphics[width=\linewidth]{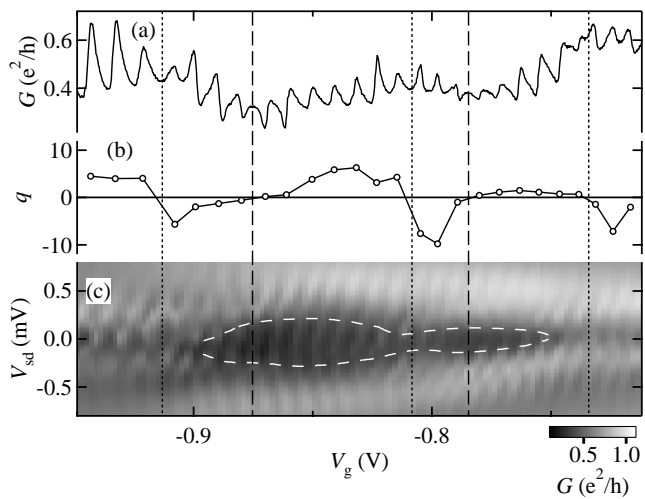}
\caption{(a) Coulomb oscillation under lower conductance condition.
Each asymmetric line shape in (a) is fitted to obtain $q$ as displayed
in (b).  Vertical broken lines indicate where the sign of $q$ changes.
(c) Gray scale plot of $G$ versus $V_{\rm g}$ and $V_{\rm sd}$.  The
small Coulomb diamonds are modulated by the SCSs as presented by the
white broken line.}
\label{Fig4}
\end{center}
\end{figure}

The crossing points can be detected more clearly when the BO is smaller.
We adjusted the coupling
strength by the side-gate voltages so as to make $|q| > 1$.  This
adjustment makes the sign change of $q$ clearer, while it sacrifices
the clarity of the diamond-like structure in the BO because the SCSs
are shrunk and the modulation of $C_{\rm eff}$ becomes weaker.

Figure~\ref{Fig4}(a) shows the Coulomb oscillation at the weaker
coupling conditions (the average conductance $\sim 0.5~e^2/h$).
The peaks show clear Fano distortion and have large enough $|q|$ for
their sign to be distinguished. In Fig.~\ref{Fig4}(b), $q$ is plotted
as a function of $V_{\rm g}$ obtained by the fitting. The $V_{\rm g}$
positions where the sign of $q$ changes are indicated by the vertical
dotted and dashed lines, and they are again placed at the peaks and
the valleys of the BO, respectively.  In Fig.~\ref{Fig4}(c), we show a
gray-scale plot of $G$ as a function of both $V_{\rm g}$ and $V_{\rm
sd}$.  Though larger diamonds are not so clear as in
Fig.~\ref{Fig1}(c), the modulation of widths of the small Coulomb
diamonds is still noticeable.  In order to see the modulation clearer,
we connect the edges of black regions that are distorted from the
diamond shape due to the Fano effect with the white broken line.  The
vertical broken lines where the sign changes go through the narrowest
points and broadest points of the region enclosed by the white line.
This adds a further support to the above discussion.

In summary, we have observed Fano interference arising from the
transmission through two energy levels inside a QD.  One of the levels
has stronger coupling to the leads and dominates the phase at the
Coulomb peak, which was confirmed from the $V_{\rm sd}$ dependence and
the analysis of the Fano resonance.  Our results support the
theoretical models that consider the existence of SCSs as the origin
of in-phase Coulomb peaks.

\acknowledgement

We thank T.~Nakanishi for helpful discussion.  This work is supported
by a Grant-in-Aid for Scientific Research from the Ministry of
Education, Culture, Sports, Science, and Technology of Japan and also
by one for Young Scientists (B) (No.~14740186) from Japan Society for
the Promotion of Science.

\end{document}